\documentclass[12pt,a4paper]{elsart}
\usepackage{latexsym,amssymb,epsfig}
\usepackage{amsmath,eepic}

\newcommand{\0}{$\phantom{0}$}

\def\0{\phantom{0}}
\renewcommand{\thefootnote}{\fnsymbol{footnote}}
\renewcommand{\thetable}{\Roman{table}} 
\begin{document}
\begin{center}
{\bf \large Transport Properties of Anisotropic Polar Fluids:}

{\bf \large 2. Dipolar Interaction}
\bigskip

G.A. Fern\'{a}ndez, J. Vrabec\footnote{To whom correspondence should be addressed, tel.:
+49-711/685-66107, fax: +49-711/685-66140, email: vrabec@itt.uni-stuttgart.de}, and H. Hasse

Institute of Thermodynamics and Thermal Process Engineering,

University of Stuttgart, D-70550 Stuttgart, Germany

\end{center}

\renewcommand{\thefootnote}{\alph{footnote}}
\baselineskip25pt

\vskip3cm Number of pages: 38

Number of tables: 1

Number of figures: 12

\clearpage
{\bf ABSTRACT}\\

Equilibrium molecular dynamics simulation and the Green-Kubo formalism
were used to calculate self-diffusion coefficient, shear viscosity, and thermal
conductivity for 38 different dipolar two-center Lennard-Jones fluids
along the bubble line and in the homogeneous
liquid. It was systematically investigated how anisotropy, i.e. elongation, and
dipole momentum influence the transport properties.
The reduced elongation $L^*$ was varied from 0 to 1 and the 
reduced squared dipole momentum was
varied depending on the elongation as follows: for spherical fluids ($L^*$=0)
from $\mu^{*2}$=0 to 20, for $L^*$=0.2 from $\mu^{*2}$=0 to 16, and for $
L^*$=0.4 and above from $\mu^{*2}$=0 to 12. 
This represents the entire range in which
parameters for real fluids are expected.
The statistical uncertainty of the reported data varies with transport property, 
for self-diffusion coefficient data the
error bars are typically lower than 3 $\%$, for shear viscosity and thermal conductivity
they are about 8 and 12 \%, respectively.

{\bf KEYWORDS:} Green-Kubo; molecular dynamics; di\-pole; self-diffusion;
shear viscosity; thermal conductivity.

\bigskip

\clearpage
\section{INTRODUCTION}

In an accompanying work \cite{fernandez7} we report on a comprehensive
investigation on
self-diffusion coefficient, shear viscosity, and thermal conductivity of
two-center Lennard-Jones plus point quadrupole (2CLJQ) fluids by equilibrium
molecular dynamics and the Green-Kubo formalism. In the present work, that
investigation is extended to two-center Lennard-Jones plus point dipole
(2CLJD) fluids using the same methodology.
These two model types are aimed at molecules with different type of polarity, i.e.
2CLJQ models can well be used for quadrupolar fluids such as CO$_2$ or Chlorine
\cite{vrabeczus}, whereas 2CLJD models are for dipolar fluids such as CO or partly fluorinated
alkanes \cite{stoll2}.

Static thermodynamic properties of the 2CLJD fluids have been extensively
investigated in the past. There are many results for vapor-liquid equilibria
\cite{saager1,saager2,dubey,muller,stoll1}, excess properties \cite{kriebel1}, or
surface tension \cite{mecke,enders}. The 2CLJD potential has been successfully
applied to modeling real fluids, yielding good results for vapor-liquid equilibria
\cite{kohler,lisal,kriebel2,stoll2}, Joule-Thomson inversion curves \cite{vrabec},
virial coefficients \cite{vega}, and recently also for shear viscosity and thermal
conductivity of refrigerants \cite{fernandez5}. On the other hand, transport
properties for this useful model class have not been explored systematically. An
exception is the work of Lee and Cummings \cite{lee} on shear viscosity using the
Stockmayer potential \cite{stockmayer}, which is a subgroup ($L^*$=0) of the 2CLJD
model fluids investigated here. The present work aims to fill this gap.

This investigation, as the accompanying one on 2CLJQ fluids, is carried out for
state points along the bubble line and in the homogeneous liquid. It covers 38
model fluids, specified each by a certain combination of elongation and dipole momentum.
Thermodynamic conditions were selected on the basis of previous work of Stoll et
al. \cite{stoll1,stoll_t} on vapor-liquid equilibria of 2CLJD fluids.
As the present work is analogous to the accompanying one on 2CLJQ
fluids \cite{fernandez7}, redundant information like the Green-Kubo
equations is omitted here.

\section{MOLECULAR MODEL}

The intermolecular interactions are represented by the two-center Lennard-Jones plus point
dipole (2CLJD) potential. The 2CLJD potential is pairwise additive and consists
out of two identical
Lennard-Jones sites a distance $L$ apart (2CLJ) plus a point dipole of momentum $\mu$ placed
in the geometric center of the molecule, oriented along the molecular axis connecting the
two Lennard-Jones (LJ) sites. It is analogous to the quadrupolar 2CLJQ potential
investigated in the accompanying work \cite{fernandez7}, only differing in the polar
interaction that is a point dipole instead of a point quadrupole. The contribution of a
point dipole is given by \cite{graygubb}

\begin{eqnarray}\label{dpoten}
u_{\rm D}=\frac{1}{4 \pi \epsilon_0} \frac{\mu^2}{\left|{\textbf{r}}^3_{ij}\right|}
(\sin \theta_i \sin \theta_j \cos \phi_{ij}- 2\cos \theta_i \cos \theta_j),
\end{eqnarray}

wherein $\textbf{r}_{ij}$ is the center-center distance vector of two molecules $i$
and $j$. $\theta_i$ is the angle between the axis of the molecule $i$ and the
center-center connection line and $\phi_{ij}$ is the azimuthal angle between the
axis of molecules $i$ and $j$. Finally, $\epsilon_0$ is the electric constant
8.854187817$\cdot$10$^{-12}$ C$^2$/(J m).

Analogously to specific 2CLJQ models, a 2CLJD model for a real substance is fully
determined by five parameters: $\sigma$, $\epsilon$, $L$, $\mu$ \cite{stoll1} and
the molecular mass $m$. In the reduced form also for this model class
only two molecular parameters remain, i.e. reduced elongation $L^*=L/ \sigma$ and
reduced squared dipole momentum $\mu^{*2}=\mu^2 /(4\pi \epsilon_0 \epsilon
\sigma^3)$. Henceforth, "squared" will be omitted in the text for brevity.

Furthermore, all results of this study were obtained and are presented in the
reduced form; the definitions are given in the accompanying work \cite{fernandez7}.
For the sake of brevity, "reduced" will be omitted
in the following.

\section{INVESTIGATED MODELS AND STATES}

In the present work, 38 different model fluids were studied, 
where each fluid is fully determined by one combination of elongation 
$L^*$ and dipole momentum $\mu^{*2}$. 
Simulations at 12 liquid state points were carried out for each model fluid.

The studied model fluids have elongations that vary from $L^*$=0,
i.e. spherical molecules, to $L^*$=1, i.e. strongly elongated dumbbell-shaped
molecules, in seven steps. For Stockmayer fluids, seven dipole momenta from
$\mu^{*2}$=0 to 20 were considered. For fluids with $L^*$=0.2, six dipole momenta
from $\mu^{*2}$=0 to 16 and for larger elongations five dipole momenta from
$\mu^{*2}$=0 to 12 were taken into account. 
The upper limit of 12 is sufficient to
describe strongly dipolar real fluids, e.g. CClF$_3$ with $\mu=1.8261$ D
($\mu^{*2}=3.4932$), CCl$_3$F with $\mu=2.7009$ D ($\mu^{*2}=3.6250$), or
CH$_2$F-CF$_3$ with $\mu=3.0214$ D ($\mu^{*2}=8.0004$). Very high dipole momenta
($\mu^{*2}$=16 and 20) were only considered for fluids with a small elongation, 
because they are realistic only for such molecules \cite{stoll2}.

Due to consistency, spherical fluids ($L^*$=0) were treated as two 
superimposed Lennard-Jones sites. To compare
with one-center Lennard-Jones results, the temperature has to be divided by 4 as
well as the dipole momentum. Also the present values of self-diffusion coefficient
$D^*$, shear viscosity $\eta^*$, and thermal conductivity $\lambda^*$ have to be
divided by 2.

To facilitate a meaningful comparison of the transport properties, 
another reduced form of the temperature $T_R$=$T^*/T_c^*$ 
and the density $\rho_R$=$\rho^* /\rho_c^*$ was used, where $T_c^*$ is the 
critical temperature and $\rho_c^*$ the critical density of the individual 
2CLJD fluid; values for $T_c^*$ and $\rho_c^*$ were taken from \cite{stoll1}.

\begin{figure}[ht]
\caption[Phase diagrams for two selected elongated 2CLJD fluids ($L^*$=0.2)
where one is non-polar and
the other strongly dipolar. Saturated densities, taken from \cite{stoll_t}, are represented by the lines
joining at the critical point depicted by $\bullet$. The investigated state points are indicated by
$\circ$ for $\mu^{*2}$=0 and by $\bigtriangleup$ for $\mu^{*2}$=12.]{} \label{fig1}
\begin{center}
\includegraphics[width=150mm,height=200mm]{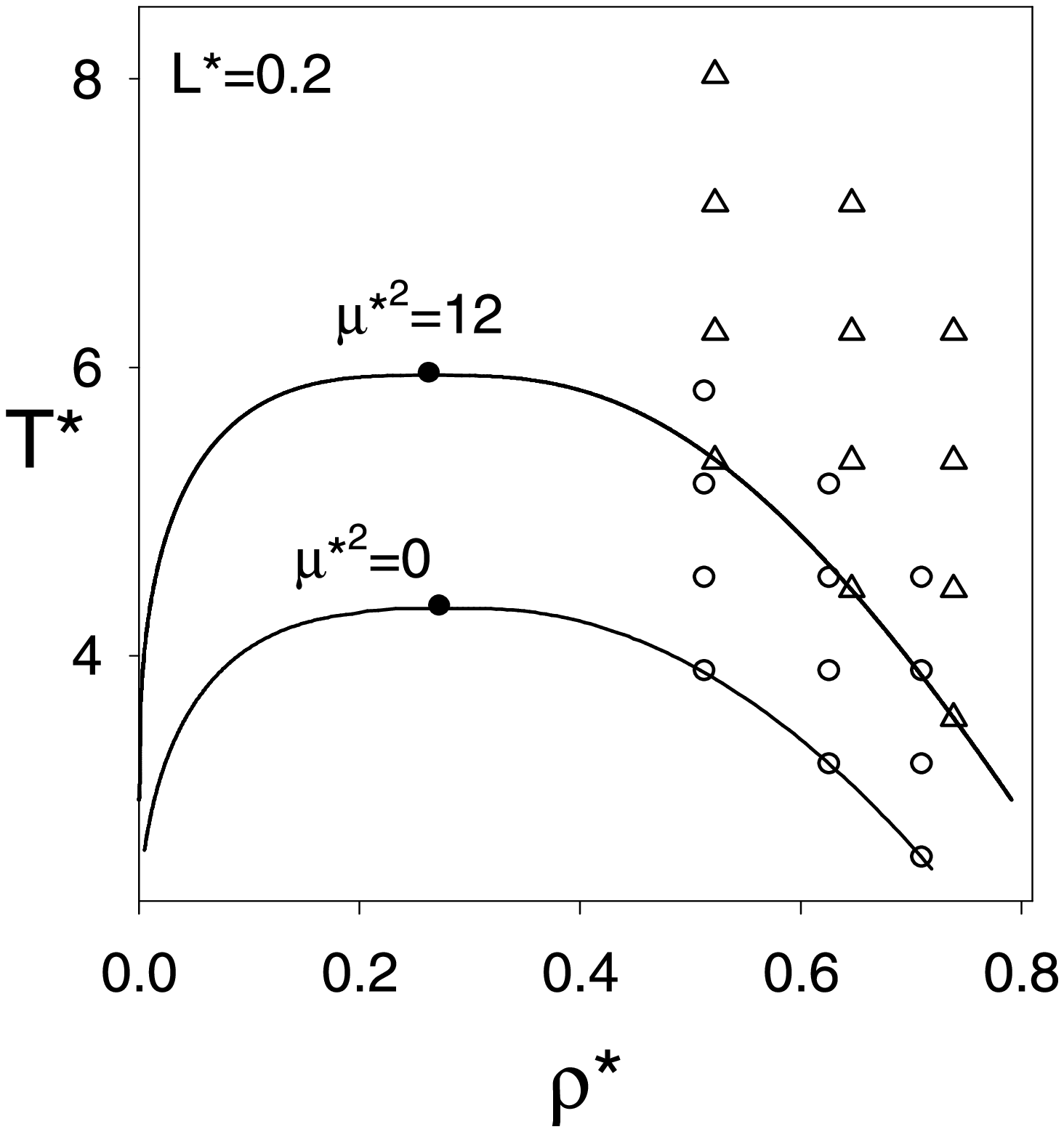}
\end{center}
\end{figure}

For each fluid, three state points along the bubble line were studied from
$T_R$=0.6 to 0.9 with temperature increments of $\Delta T_R$=0.15. In addition,
another nine state points were studied in the homogeneous liquid on three
isochores, starting from these three bubble points with temperature increments of
$\Delta T_R$=0.15, cf. Fig. \ref{fig1}. 
This grid is less dense than the one used in the accompanying work on 
quadrupolar fluids \cite{fernandez7}, however, as more model fluids are
regarded here, both present roughly the same number of data points.

\begin{figure}[ht]
\caption[Saturated liquid density of Stockmayer fluids ($L^*$=0) for three reduced
temperatures ($T_R$=0.6, 0.9, and 1) as functions of the dipole momentum. The solid
lines indicate the overall correlation in \cite{stoll1}, whereas the
symbols are from correlations fitted to each model fluid's vapor-liquid
equilibrium individually.]{}
\label{fig2}
\begin{center}
\includegraphics[width=150mm,height=200mm]{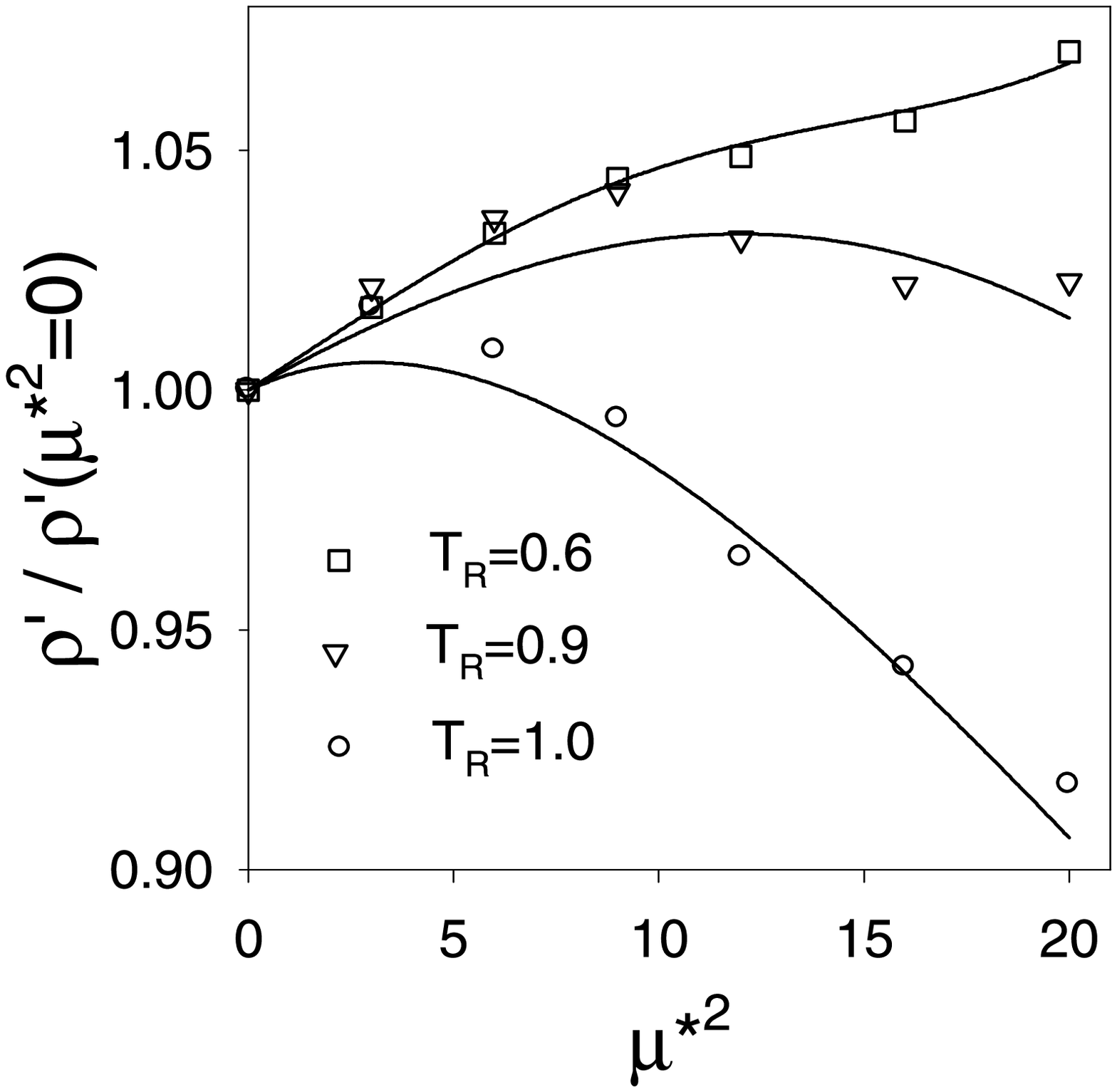}
\end{center}
\end{figure}

For quadrupolar fluids critical and bubble densities in terms of $\rho^*$ 
increase monotonously with increasing quadrupole momentum \cite{stoll1},
but dipolar fluids behave different \cite{stoll2}. Fig. \ref{fig2} illustrates the
variation of the bubble density, for three reduced temperatures as a function of
the dipole momentum for Stockmayer fluids ($L^*$=0). As can be seen, the critical
density, i.e. bubble density at $T_R$=1, increases up to $\mu^{*2}\approx$ 3,
but decreases for higher dipole momenta. At $T_R$=0.9 and below they are larger
than those of non-polar fluids in all cases, where for this temperature a
significant maximum of the bubble density at intermediate dipole momenta is
present. These tendencies are also found for elongated fluids. The influence of
elongation on phase behavior does not differ from that of quadrupolar fluids
\cite{fernandez7}, i.e. bubble densities decrease
monotonously with increasing elongation \cite{stoll1}.

\section{SIMULATION DETAILS}
As many of the considerations and technical details of the simulations are the
same as those used in the accompanying work \cite{fernandez7}, we proceed 
describing only the differences. Here, the significant long range contributions due to the dipolar
interaction were treated with the reaction field technique
\cite{barker,neumann}. Simulations were carried out in the canonical 
ensemble using the modified
equations of motion with a time step of $\Delta t \sqrt{\epsilon /
m}/\sigma=0.001$, and the Nose-Hoover thermostat \cite{nose,frenkel} with a
thermal inertial parameter of 10 in reduced units. Statistical uncertainties
of the transport coefficients were estimated using the standard deviation of three
independent simulations runs with $5~000$ independent autocorrelation functions.
Independence between correlation functions was achieved with a time span of 0.2
in reduced units between consecutive correlation functions.
The convergency issues of self-diffusion
coefficient \cite{alder3} and shear viscosity
\cite{alder2,fernandez3,levesque1,schoen,erpenbeck2} are discussed in the
accompanying work \cite{fernandez7}. The integrals usually converge within
their statistical uncertainties at about 1 in reduced units, 
nevertheless an integration time of $\Delta t^*$=2 was used.

\section{RESULTS}
In this section, the simulation results for the transport coefficients are presented.
Numerical data for self-diffusion coefficient, shear viscosity, and thermal conductivity
are given in Table \ref{tab1a} for spherical fluids ($L^*$=0) with dipole momenta from $\mu^{*2}$=0 to
20, for slightly elongated fluids ($L^*$=0.2) with dipole momenta from $\mu^{*2}$=0 to 16, and
for elongated fluids from $L^*$=0.4 to 1 with dipole momenta ranging from $\mu^{*2}$=0 to
12. All data in Table \ref{tab1a} correspond to state points along the bubble line
for reduced temperatures of $T_R=$0.6, 0.75, and 0.9. The
complete data set, with nine additional state points in the liquid region for each fluid, is
available in \cite{fernandezt} and partially included in Figs. \ref{fig6}, \ref{fig7},
\ref{fig9}, \ref{fig10}, \ref{fig12}, and \ref{fig13}. The effects of elongation, dipole
momentum, temperature, and density are discussed in the following for each transport
coefficient separately.

The accuracy of the calculated transport properties decreases in the sequence self-diffusion
coefficient, shear viscosity, thermal conductivity. The high accuracy of the
self-diffusion coefficient, with error bars lower than 3 \%, is due to its individual nature
\cite{hansen}. Shear viscosity and thermal conductivity show larger uncertainties,
that are around 8 and 12 \%, respectively.
In most simulations of the present work the autocorrelation functions of thermal
conductivity decay faster than those for
shear viscosity, but fluctuate more.

Other factors that influence the accuracy of the reported data are elongation and dipole
momentum. In particular, at low temperatures, for fluids with large anisotropy and strong
dipole momentum, the transport coefficients show larger simulation uncertainties.

In the following, the results are discussed for ten selected fluids, covering the 
whole range of the two molecular parameters, from spherical ($L^*$=0) over elongated 
($L^*$=0.505) to  strongly elongated ($L^*$=1.0) fluids with varying dipole 
momentum of $\mu^{*2}$=0, 6, 12, and 20. 
A subset of six fluids is taken in some cases only due to graphical reasons.

\subsection{Self-diffusion coefficient}
\begin{figure}[ht]
\caption[Self-diffusion coefficient 
of spherical ($L^*$=0, empty symbols), 
elongated ($L^*$=0.505, grey symbols), 
and strongly elongated ($L^*$=1, full symbols) 
2CLJD fluids over reduced density along bubble lines. 
Reduced temperatures vary from $T_R$=0.6 to 0.9.
Lines are guides for the eye.]{} \label{fig4}
\begin{center}
\includegraphics[width=150mm,height=200mm]{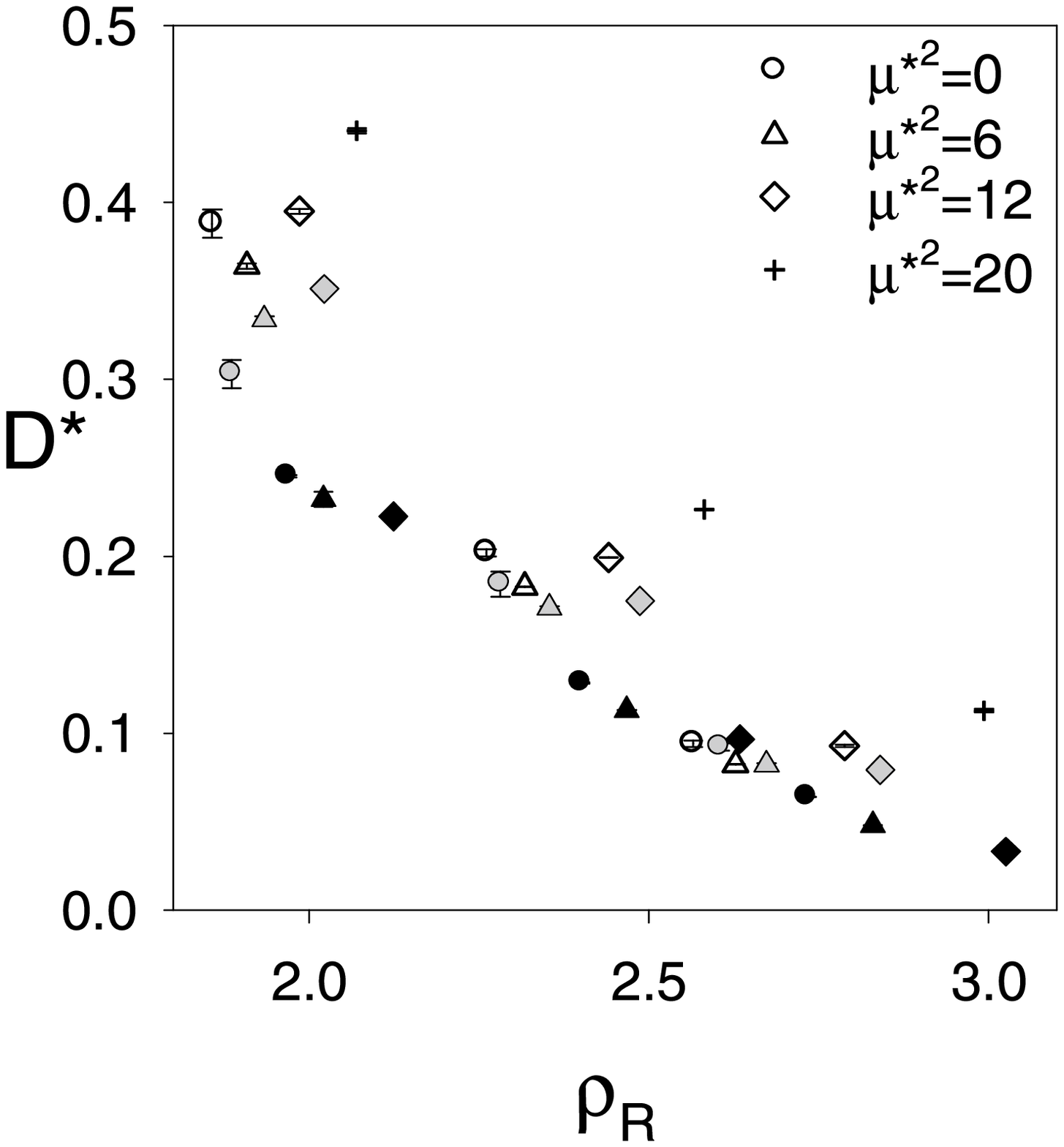}
\end{center}
\end{figure}

\begin{figure}[ht]
\caption[Self-diffusion coefficient 
of spherical ($L^*$=0, empty symbols), 
elongated ($L^*$=0.505, grey symbols), 
and strongly elongated ($L^*$=1, full symbols)  
2CLJD fluids over number density along bubble lines. 
Reduced temperatures vary from $T_R$=0.6 to 0.9.
Lines are guides for the eye.]{} \label{fig5}
\begin{center}
\includegraphics[width=150mm,height=200mm]{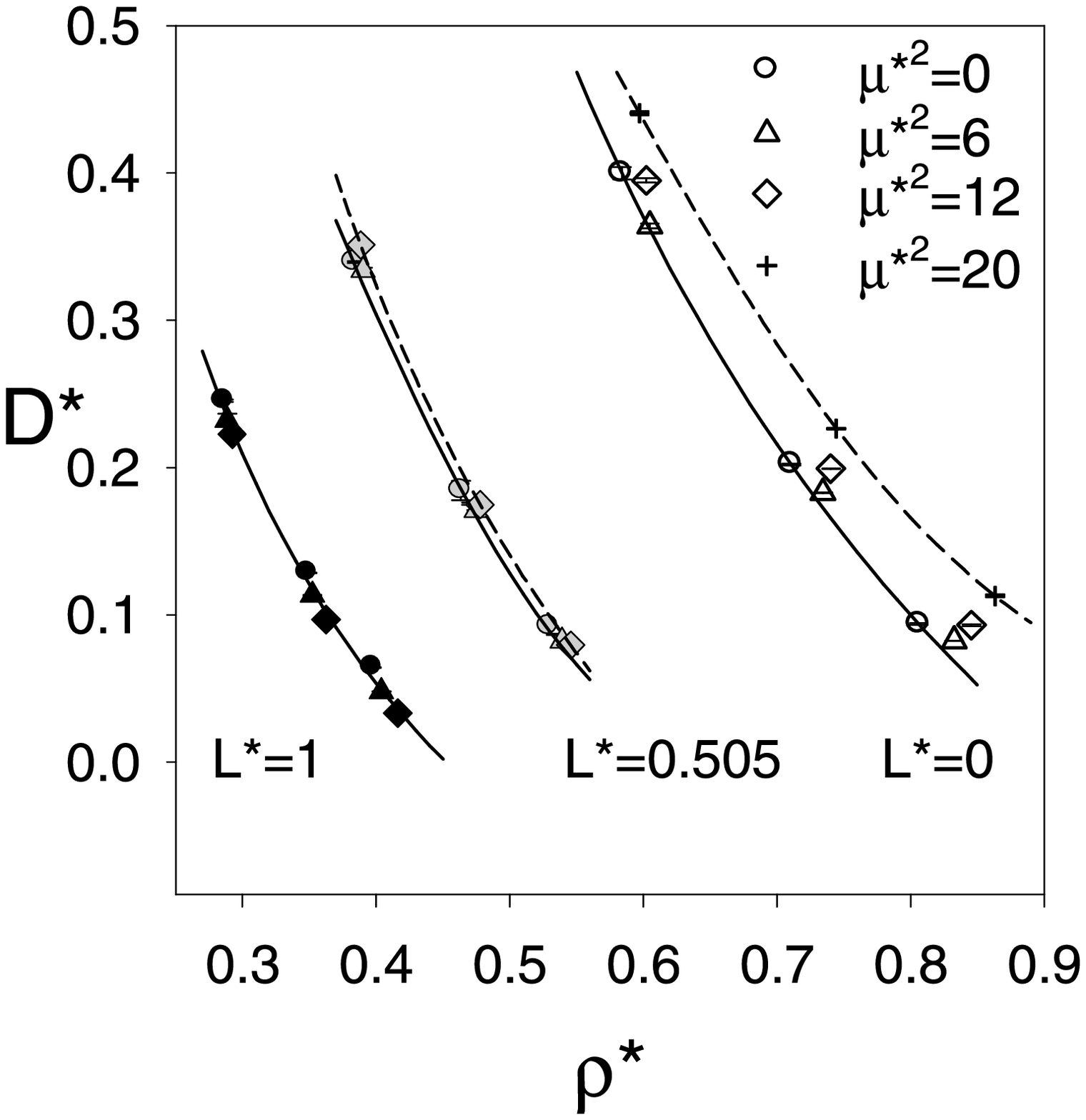}
\end{center}
\end{figure}

\begin{figure}[ht]
\caption[Self-diffusion coefficient 
of spherical ($L^*$=0, empty symbols), 
elongated ($L^*$=0.505, grey symbols), 
and strongly elongated ($L^*$=1, full symbols) 
2CLJD fluids over number density 
in the homogeneous liquid at $T_R$=0.9.
Lines are guides for the eye.]{} \label{fig6}
\begin{center}
\includegraphics[width=150mm,height=200mm]{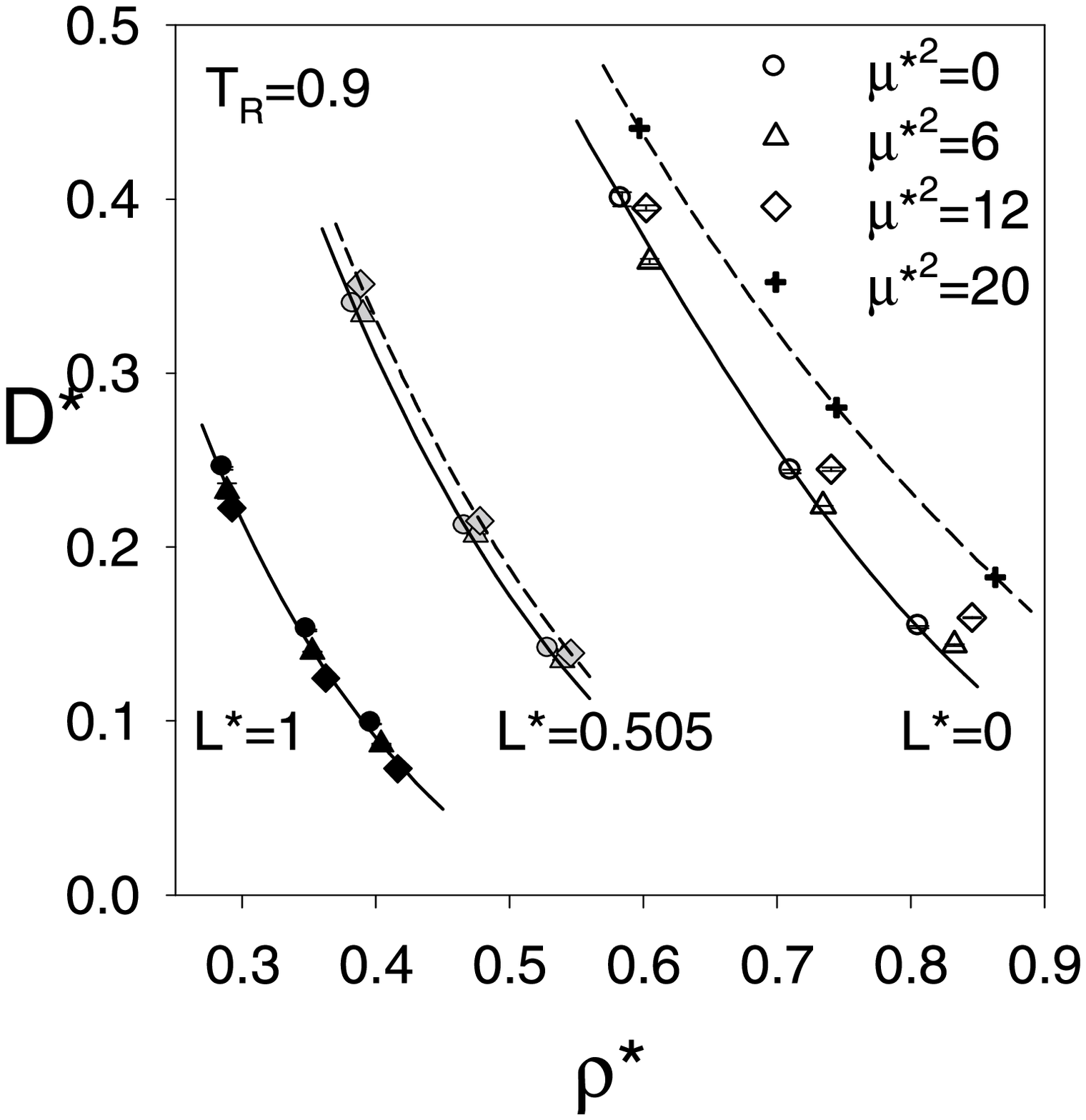}
\end{center}
\end{figure}

\begin{figure}[ht]
\caption[Self-diffusion coefficient of spherical ($L^*$=0, empty symbols) and
strongly elongated ($L^*$=1, full
symbols) 2CLJD fluids over reduced temperature 
in the homogeneous liquid along different isochores. 
$\circ$: $\rho^*$=0.8062,
$\vartriangle$: $\rho^*$=0.8327, $\lozenge$: $\rho^*$=0.8456.
Lines are guides for the eye.]{} \label{fig7}
\begin{center}
\includegraphics[width=150mm,height=200mm]{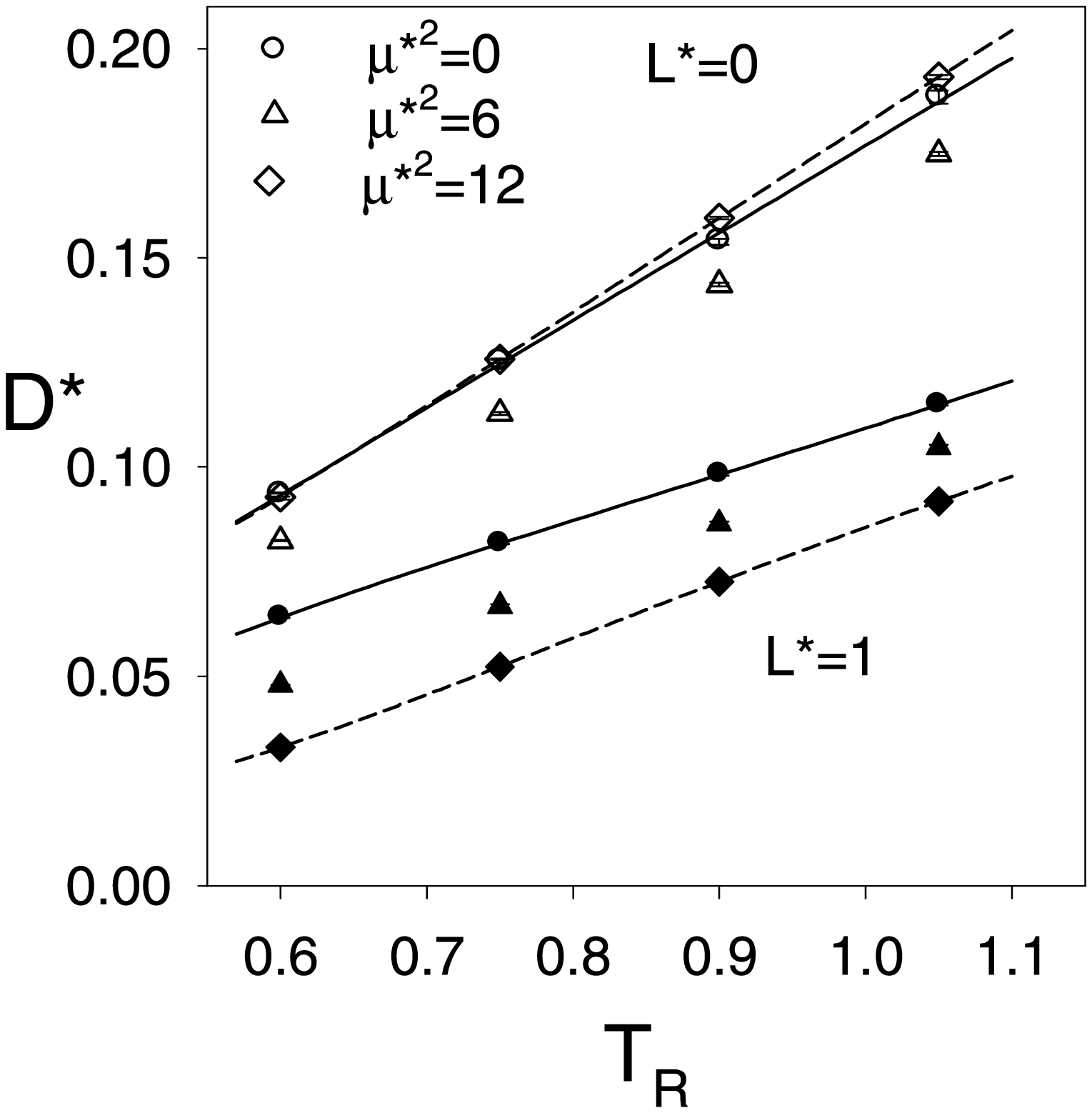}
\end{center}
\end{figure}

Figs. \ref{fig4} and \ref{fig5} illustrate the self-diffusion coefficient along
the bubble line for ten selected fluids. 
The results can either be discussed in terms of reduced density $\rho_R$ 
as in Fig. \ref{fig4} or in terms of number density $\rho^*$ as in Fig. \ref{fig5}.
From Fig. \ref{fig4} it can be seen that the regarded range of reduced density is similar 
for all fluids, but significant deviations from the principle of corresponding states 
are present also for the density. 
At constant $T_R$, it can be discerned that the self-diffusion
coefficient always decreases with increasing elongation.
The dipole, however, can either decrease or increase the self-diffusion coefficient. 
A better visibility of the data (which is even more needed for the less accurate properties 
shear viscosity and thermal conductivity) is obtained when plotted over number density in 
Fig. \ref{fig5}. Therefore, this graphical representation is preferred in the following.

As Fig. \ref{fig5} shows, $D^*$ 
decreases with increasing number density along the bubble line.
Self-diffusion coefficients of fully elongated
fluids lie roughly along the same line as observed for quadrupolar fluids
\cite{fernandez7}. To analyze the effect of the dipole momentum, it is helpful to
compare the bubble densities of Stockmayer fluids in Fig. \ref{fig2}, 
where the influence is most
visible, with the corresponding self-diffusion coefficients in Fig.
\ref{fig5}. The self-diffusion coefficient shows a minimum for Stockmayer fluids
with $\mu^{*2}\approx$ 6 at all three temperatures. The bubble densities of
Stockmayer fluids also show a peculiar behavior: maxima at $T_R$=0.9 and 1 
and a point of inflexion at $T_R$=0.6. 
As the self-diffusion coefficient of non-polar fluids decreases with increasing density, 
it can be concluded that the isolated effect of the dipole is to increase the
self-diffusion coefficient. 

Fig. \ref{fig6} shows the dependence of $D^*$ on number density in the homogeneous liquid
region at a constant reduced
temperature of $T_R$=0.9. Note that the density range is the same as in Fig.
\ref{fig5}. Along this isotherm $D^*$ decreases slightly hyperbolic with
increasing density, resembling the behavior of $D^*$ along bubble lines for a given
elongation. Comparing $D^*$ along bubble lines with isothermal data for the same
density variation, it is found that density is the dominant variable, being
responsible for about 80 \% of the variation in $D^*$.

Fig. \ref{fig7} shows the dependence of the self-diffusion coefficient on reduced
temperature for a subset of six selected fluids at different isochores.
The isochores correspond to bubble densities at the reduced temperature
$T_R$=0.6, cf. Fig. \ref{fig1}, which have similar values in terms of
$\rho_R$. It can be seen that the self-diffusion coefficient
decreases monotonously with increasing dipole momentum for elongated fluids,
whereas for Stockmayer fluids, $D^*$ again exhibits a minimum at intermediate
dipole momenta. Along an isochore, the self-diffusion coefficient increases
linearly with increasing temperature. The gradients with respect to reduced
temperature are almost constant for a given elongation but less pronounced for
more elongated fluids. Such a behavior was also
found in the accompanying work on quadrupoles \cite{fernandez7}.

\subsection{Shear viscosity}
\begin{figure}[ht]
\caption[Shear viscosity 
of spherical ($L^*$=0, empty symbols), 
elongated ($L^*$=0.505, grey symbols), 
and strongly elongated ($L^*$=1, full symbols) 
2CLJD fluids over number density along bubble lines. 
Reduced temperatures vary from $T_R$=0.6 to 0.9.
Lines are guides for the eye.]{} \label{fig8}
\begin{center}
\includegraphics[width=150mm,height=200mm]{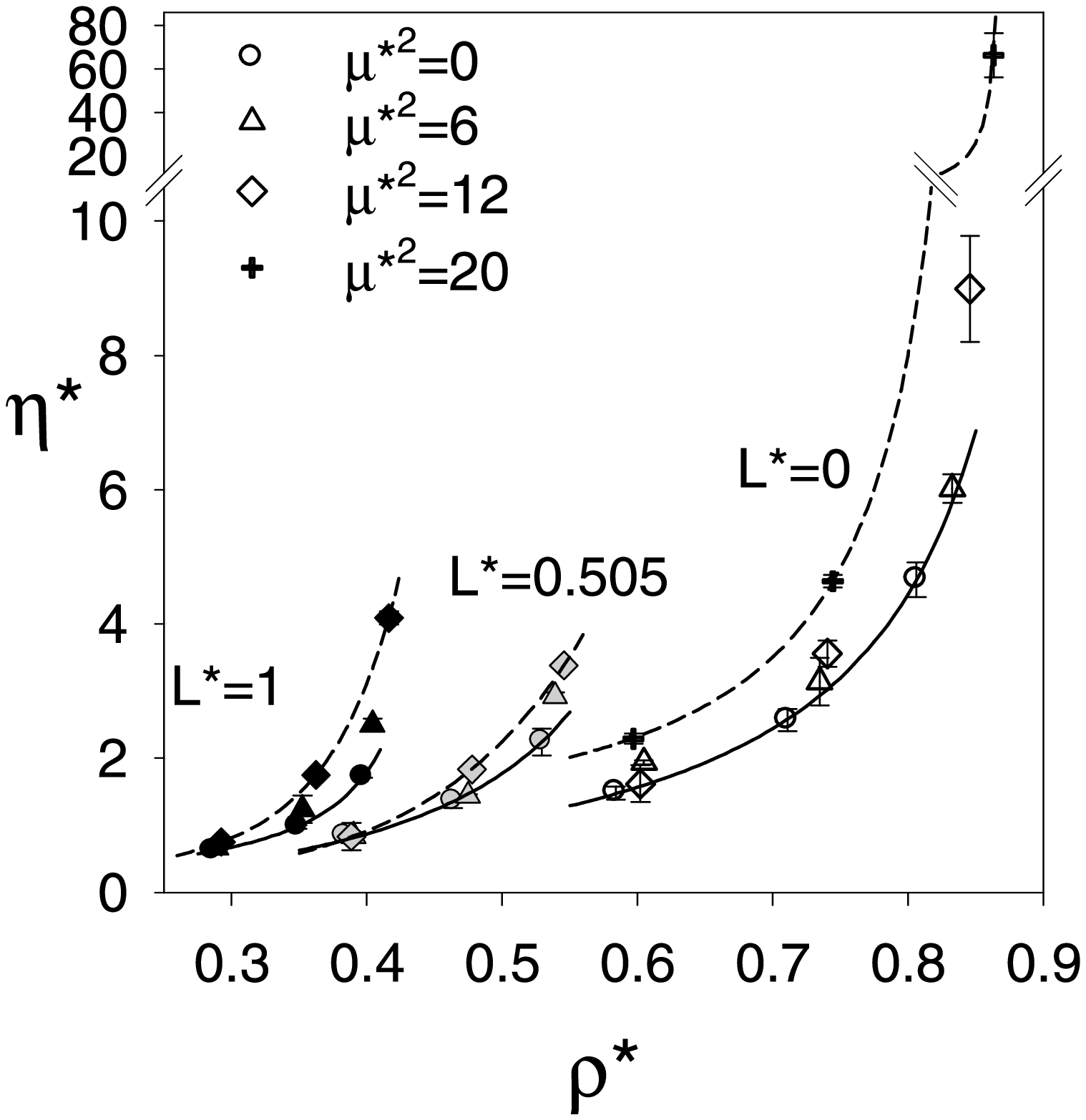}
\end{center}
\end{figure}

\begin{figure}[ht]
\caption[Shear viscosity 
of spherical ($L^*$=0, empty symbols), 
elongated ($L^*$=0.505, grey symbols), 
and strongly elongated ($L^*$=1, full symbols) 
2CLJD fluids over number density 
in the homogeneous liquid
at $T_R$=0.9.
Lines are guides for the eye.]{} \label{fig9}
\begin{center}
\includegraphics[width=150mm,height=200mm]{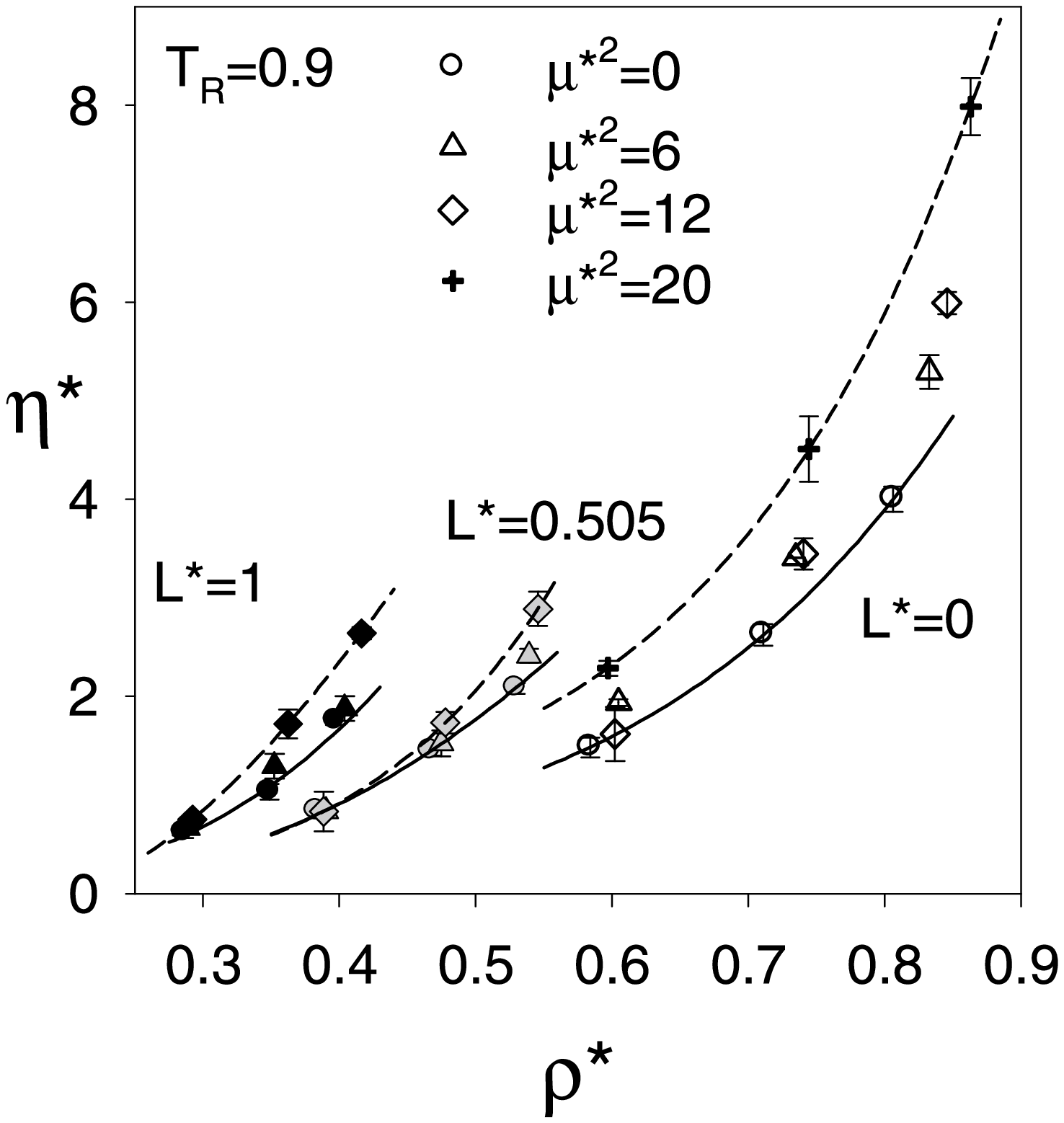}
\end{center}
\end{figure}

\begin{figure}[ht]
\caption[Shear viscosity of spherical ($L^*$=0, empty symbols) 
and strongly elongated ($L^*$=1, full
symbols) 2CLJD fluids over reduced temperature 
in the homogeneous liquid along different isochores. $\circ$: $\rho^*$=0.8062,
$\vartriangle$: $\rho^*$=0.8327, $\lozenge$: $\rho^*$=0.8456.
Lines are guides for the eye.]{} \label{fig10}
\begin{center}
\includegraphics[width=150mm,height=200mm]{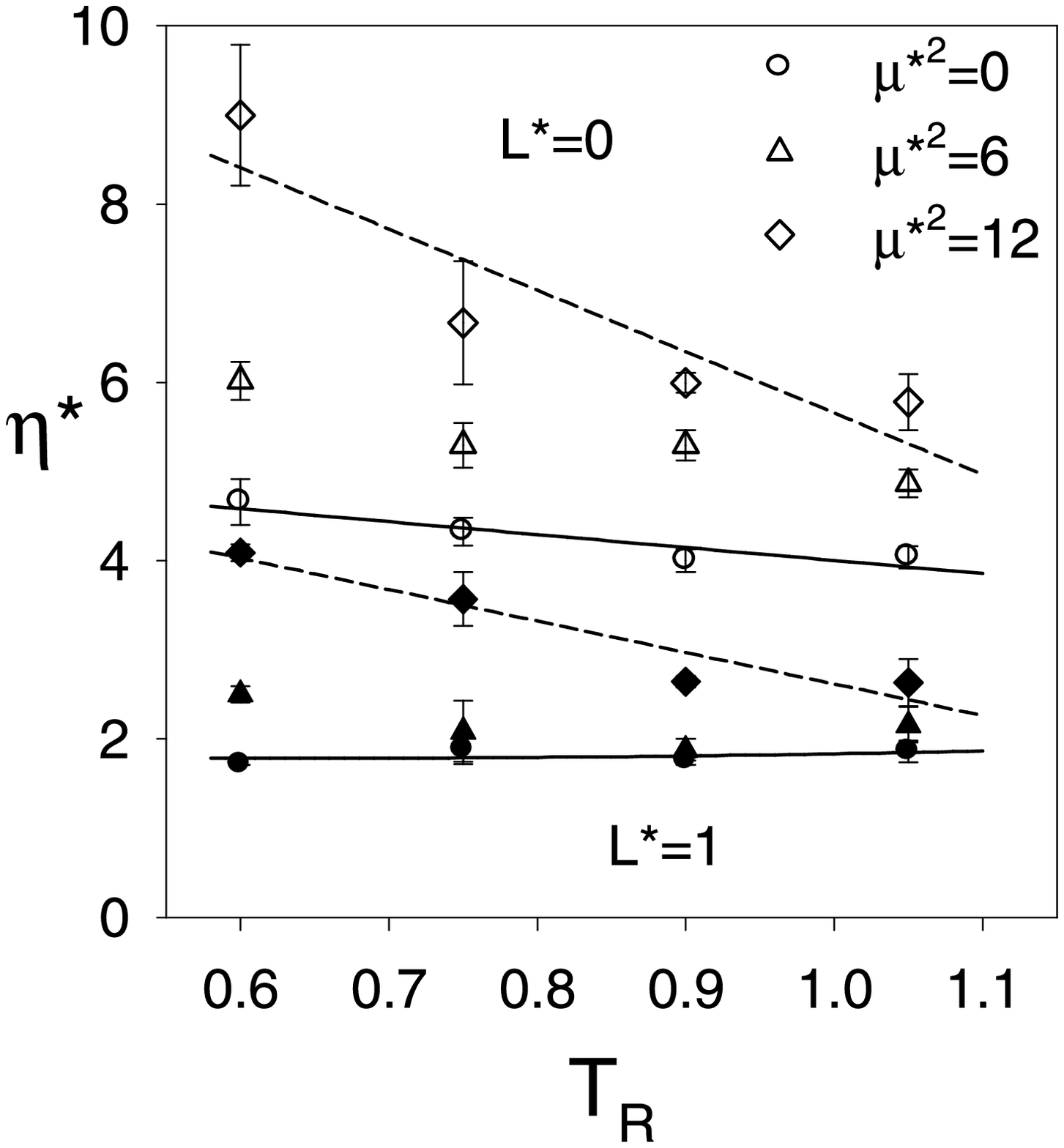}
\end{center}
\end{figure}

Fig. \ref{fig8} illustrates the shear viscosity along the bubble line 
for the ten selected fluids.
In contrast to quadrupolar fluids, shear viscosity results for fluids with a 
given elongation, but different polar momentum, are not along a single line. 
At constant $T_R$, shear viscosity increases considerably 
with increasing dipole momentum. 
In accordance to quadrupolar fluids \cite{fernandez7}, more elongated molecules 
have generally a lower shear viscosity along the bubble line.
The extremely high values of shear viscosity found for high dipole momenta 
($\mu^{*2}$= 20) at low temperatures are remarkable. 

Fig. \ref{fig9} illustrates the isolated effect of density on
shear viscosity for the same ten fluids in the homogeneous liquid region 
at $T_R$=0.9. Comparing the variation of
$\eta^*$ along bubble lines and along isotherms in the same way as for $D^*$, it
is found that the density effect is responsible for about 80 \% of the increase
of $\eta^*$ along the bubble line. Exceptions are the shear viscosities of
strongly dipolar Stockmayer fluids with $\mu^{*2}$=16 and 20. These are
very sensitive to temperature changes, where the contribution of density is only
about 10 \% for $\mu^{*2}$=20.

Fig. \ref{fig10} shows the dependence of shear viscosity on reduced temperature
for a subset of six fluids along different isochores 
with similar values in terms of $\rho_R$. As expected, the
shear viscosity decreases with increasing temperature. Non-polar fluids show little
temperature dependence. On the other hand, strongly dipolar fluids, with an about
twofold higher shear viscosity than non-polar fluids in the cold liquid, are more
sensitive to temperature exhibiting larger gradients.

\subsection{Thermal conductivity}
\begin{figure}[ht]
\caption[Thermal conductivity 
of spherical ($L^*$=0, empty symbols), 
elongated ($L^*$=0.505, grey symbols), 
and strongly elongated ($L^*$=1, full symbols)  
2CLJD fluids over number density along bubble lines. 
Reduced temperatures vary from $T_R$=0.6 to 0.9.
Lines are guides for the eye.]{} \label{fig11}
\begin{center}
\includegraphics[width=150mm,height=200mm]{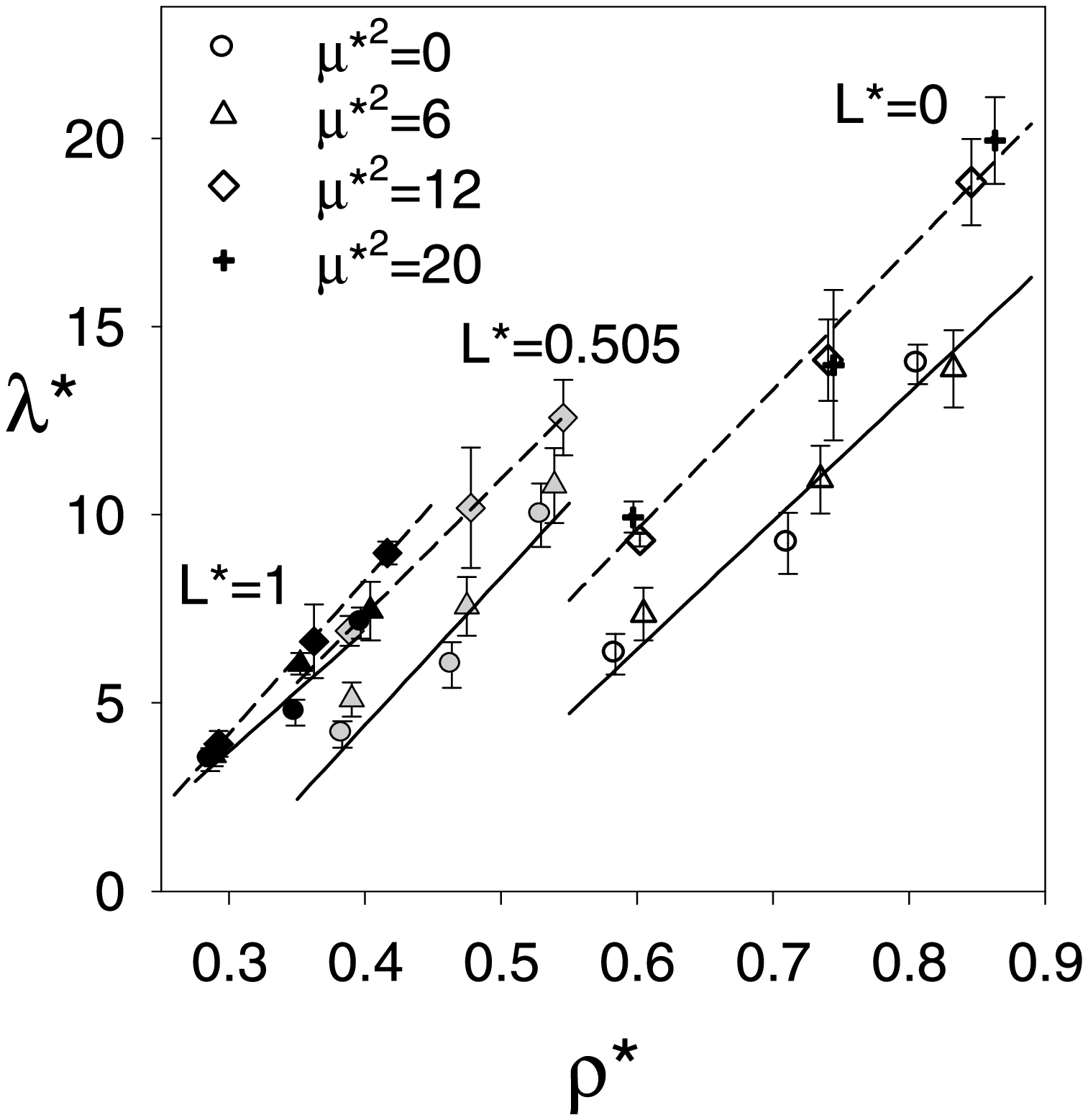}
\end{center}
\end{figure}

\begin{figure}[ht]
\caption[Thermal conductivity 
of spherical ($L^*$=0, empty symbols), 
elongated ($L^*$=0.505, grey symbols), 
and strongly elongated ($L^*$=1, full symbols) 
2CLJD fluids over number density 
in the homogeneous liquid at $T_R$=0.9.
Lines are guides for the eye.]{} \label{fig12}
\begin{center}
\includegraphics[width=150mm,height=200mm]{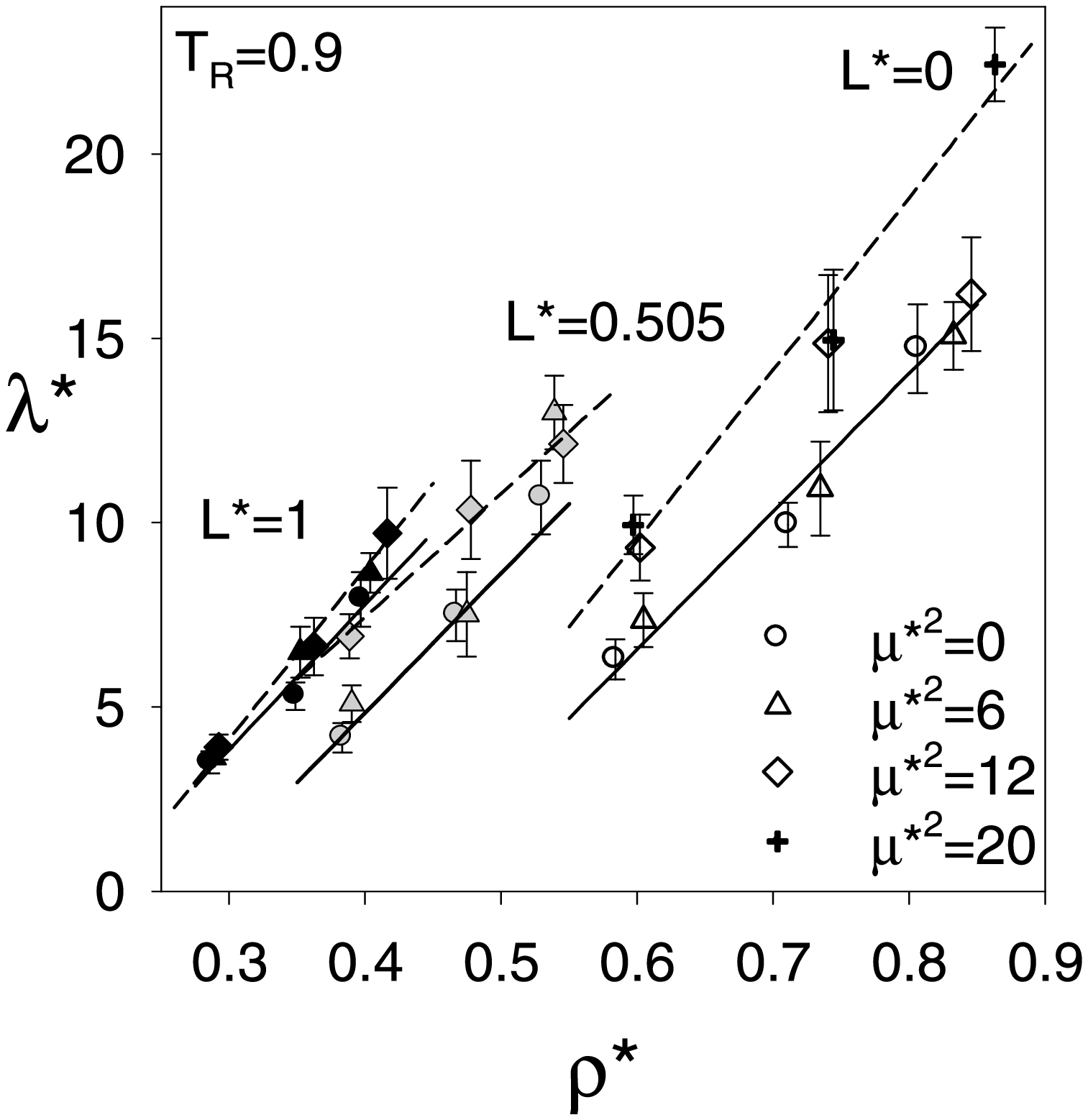}
\end{center}
\end{figure}

\begin{figure}[ht]
\caption[Thermal conductivity of spherical ($L^*$=0, empty symbols) and strongly
elongated ($L^*$=1, full
symbols) 2CLJD fluids over reduced temperature 
in the homogeneous liquid along different isochores. 
$\circ$: $\rho^*$=0.8062,
$\vartriangle$: $\rho^*$=0.8327, $\lozenge$: $\rho^*$=0.8456. Lines are guides for the eye.]{}
\label{fig13}
\begin{center}
\includegraphics[width=150mm,height=200mm]{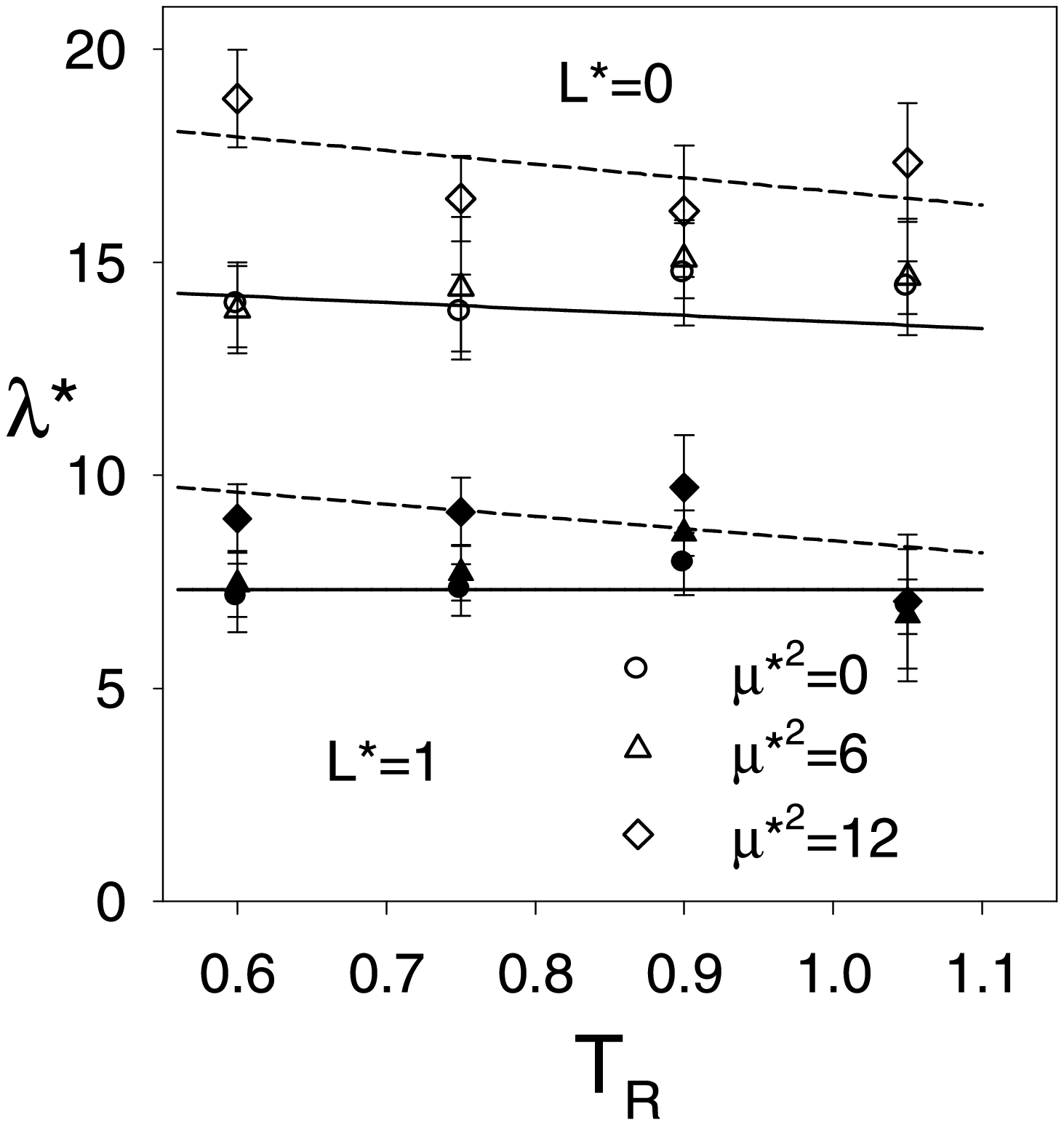}
\end{center}
\end{figure}

Fig. \ref{fig11} illustrates the thermal conductivity along the bubble line. 
The thermal conductivity increases with increasing density 
and more elongated molecules have lower thermal conductivities, 
similar to quadrupolar fluids \cite{fernandez7}. 
In contrast to quadrupolar fluids, the data for a constant elongation 
do not lie on a single line here.
It can best be seen for Stockmayer fluids that $\lambda^*$ increases 
strongly with increasing dipole momentum at constant $T_R$.

Fig. \ref{fig12} shows the density dependence of the same ten fluids 
in the homogeneous liquid at $T_R$=0.9. 
As can be seen, the curves resemble those along bubble
lines, cf. Fig. \ref{fig11}, demostrating the dominant density effect. 
Similar results were found for quadrupolar fluids \cite{fernandez7}.

Fig. \ref{fig13} shows the weak
temperature dependence of $\lambda^*$ through isochoric data for a subset of six fluids.
Taking the statistical uncertainties and the scatter into account, hardly
any trend can be discerned. This is also in agreement to the findings for quadrupolar
fluids \cite{fernandez7}.

\section{CONCLUSION}
Equilibrium molecular dynamics simulation and the Green-Kubo
formalism were used to calculate the self-diffusion coefficient, shear viscosity,
and thermal conductivity for 38 different anisotropic and dipolar model fluids. 
A comprehensive data set was obtained for each fluid and property that covers a
substantial part of the liquid state.
The statistical uncertainty of the reported data varies according
to transport property. For self-diffusion coefficient data it is lower than 3 \%,
for shear viscosity and thermal conductivity it is around 8 and 12 \%, respectively.

It is found that both anisotropy and dipole momentum significantly influence all
transport properties along the bubble line. An increasing elongation always leads to
lower values, whereas an increasing dipole momentum usually yields higher values.
The predominant part of these variations is due to the considerable variations 
in number density along the bubble line caused by $L^*$ and $\mu^{*2}$.
However, peculiar extrema are found, e.g., for the self-diffusion coefficient of 
fluids with small anisotropy. 

Density is the most important thermodynamic variable, however, compared to the 
findings for quadrupolar fluids \cite{fernandez7}, it is less dominating and the
influence of the dipole is much stronger.

Temperature influences all transport properties less than density. As expected,
for higher temperatures the self-diffusion coefficient increases, the shear
viscosity decreases, and for the thermal conductivity hardly any variation can
be discerned.

\vskip1cm

\noindent
{\bf List of symbols}

\begin{tabular}{ll}
$D$ & self-diffusion coefficient \\
$i$ & molecule index \\
$j$ & molecule index \\
$L$ & molecular elongation \\
$m$ & molecular mass \\
$t$ & time \\
$T$ & temperature \\
$u$ & pair potential \\
$\Delta$ & increment \\
$\Delta t$ & integration time step\\
$\epsilon$ & Lennard-Jones energy parameter \\
$\epsilon_0$ & Electric constant \\
$\eta$ & shear viscosity \\
\end{tabular}

\begin{tabular}{ll}
$\theta$  & angle of nutation \\
$\lambda$ & thermal conductivity \\
$\mu$     & molecular dipole momentum \\
$\rho$    & density \\
$\sigma$  & Lennard-Jones size parameter \\
$\phi$    & azimuthal angle \\
\end{tabular}

\noindent
\textbf{Vector properties} \\[0.1cm]
\begin{tabular}{ll}
$\textbf{r}$ & distance vector \\
\end{tabular}

\noindent
\textbf{Subscript} \\[0.1cm]
\begin{tabular}{ll}
$c$ & property at critical point \\
D & point dipole \\
$R$ & property reduced by critical value \\
\end{tabular}

\noindent
\textbf{Superscript} \\[0.1cm]
\begin{tabular}{ll}
*   & property reduced by molecular parameters \\
\end{tabular}

\vskip1cm

\noindent
{\bf Acknowledgement}

We gratefully acknowledge financial support from Deutscher Akademischer Austauschdienst
(DAAD). 

\clearpage

\clearpage
\begin{table}[t]
\noindent \caption{Transport coefficients along bubble lines for 38 2CLJD fluids with
different elongations $L^*$ and dipole momentum $\mu^{*2}$. The numbers in parentheses denote
the uncertainty in the last digits.} \label{tab1a}
\bigskip
\begin{center}
\begin{tabular}
{lcc|ccc} \hline\hline
$L^*$=0 & $T^*$ & $\rho^*$ & $D^*$ & $\eta^*$ & $\lambda^*$  \\
\hline\hline

$\mu^{*2}$=0    & 3.139 &  0.8062 &  0.094(2)\0 & \04.65(25) & 13.4(6)\0\0   \\
                & 3.925 &  0.7108 &  0.202(2)\0 & \02.56(16) & \09.24(81)  \\
                & 4.708 &  0.5838 &  0.388(8)\0 & \01.48(10) & \06.29(54)  \\
\hline
$\mu^{*2}$=3    & 3.288 &  0.8202 & 0.083(1)\0  & \05.46(8)\0  & 13.04(89) \\
                & 4.110 &  0.7245 & 0.184(4)\0  & \02.79(39)   & 10.12(90) \\
                & 4.932 &  0.5966 & 0.372(7)\0  & \01.74(19)   & \07.26(73)  \\
\hline
$\mu^{*2}$=6    & 3.597 &  0.8327 & 0.082(2)\0  & \06.02(21)   & 13.9(10)\0  \\
                & 4.494 &  0.7345 & 0.183(4)\0  & \03.14(35)   & 10.93(90) \\
                & 5.395 &  0.6049 & 0.364(7)\0  & \01.94(3)\0  & \07.36(70)  \\
\hline
$\mu^{*2}$=9    & 3.954 &  0.8419 & 0.084(2)\0  & \07.44(72)   & 16.41(31) \\
                & 4.943 &  0.7407 & 0.187(4)\0  & \03.41(19)   & 13.6(10)\0  \\
                & 5.931 &  0.6082 & 0.370(7)\0  & \01.95(7)\0  & \07.01(70)   \\
\hline
$\mu^{*2}$=12   & 4.378 &  0.8456 & 0.093(2)\0  & \08.99(79) & 18.8(12)\0  \\
                & 5.470 &  0.7404 & 0.199(4)\0  & \03.56(20) & 14.1(11)\0  \\
                & 6.564 &  0.6024 & 0.395(8)\0  & \01.62(28) & \09.32(17)  \\
\hline
$\mu^{*2}$=16   & 4.956 &  0.8516 & 0.104(2)\0  & 18.8(16)\0 & 20.7(21)\0 \\
                & 6.192 &  0.7400 & 0.215(4)\0  & \03.75(23) & 15.8(18)\0 \\
                & 7.429 &  0.5968 & 0.426(9)\0  & \01.93(7)\0  & \08.91(88) \\
\hline
$\mu^{*2}$=20   & 5.505 &  0.8632 & 0.113(2)\0  & 66.4(10)\0 & 19.9(12)\0 \\
                & 6.878 &  0.7445 & 0.226(5)\0  & \04.64(1)\0  & 14.0(20)\0 \\
                & 8.254 &  0.5972 & 0.441(9)\0  & \02.29(8)\0  & \09.93(42) \\

\hline
\hline
\end{tabular}
\end{center}
\end{table}

\setcounter{table}{0}
\begin{table}[t]
\noindent \caption{Continued.}
\bigskip
\begin{center}
\begin{tabular}
{lcc|ccc} \hline\hline
$L^*$=0.2 & $T^*$ & $\rho^*$ & $D^*$ & $\eta^*$ & $\lambda^*$  \\
\hline\hline
$\mu^{*2}$=0 & 2.590 &  0.7114 & 0.088(1)  &  3.64(16)   & 11.18(62) \\
             & 3.237 &  0.6275 & 0.190(1)  &  2.19(7)\0  & \08.7(12)\0 \\
             & 3.884 &  0.5144 & 0.360(8)  &  1.20(4)\0  & \05.66(35) \\
\hline
$\mu^{*2}$=3 & 2.715 &  0.7220 & 0.086(1)  &  4.21(42)   & 12.09(95) \\
             & 3.393 &  0.6367 & 0.189(1)  &  2.19(6)\0  & 10.42(60) \\
             & 4.072 &  0.5244 & 0.372(2)  &  1.34(1)\0  & \08.30(77) \\
\hline
$\mu^{*2}$=6 & 2.952 &  0.7307 & 0.091(1)  &  4.27(50)   & 17.7(11)\0 \\
             & 3.691 &  0.6439 & 0.197(1)  &  2.27(18)   & 15.73(86) \\
             & 4.428 &  0.5306 & 0.389(1)  &  1.34(4)\0  & 11.3(19)\0 \\
\hline
$\mu^{*2}$=9 & 3.233 &  0.7370 & 0.098(1)  &  4.22(1)\0  & 26.6(16)\0 \\
             & 4.040 &  0.6481 & 0.211(1)  &  2.42(5)\0  & 22.0(14)\0 \\
             & 4.848 &  0.5315 & 0.415(2)  &  1.20(3)\0  & 11.9(19)\0  \\
\hline
$\mu^{*2}$=12 & 3.572 &  0.7388 & 0.110(1) &  4.32(43)   & 29.4(21)\0    \\
              & 4.464 &  0.6463 & 0.235(1) &  2.55(34)   & 24.7(31)\0    \\
              & 5.355 &  0.5226 & 0.466(1) &  1.38(1)\0  & 11.5(15)\0    \\
\hline
$\mu^{*2}$=16 & 3.998 &  0.7438 & 0.123(1) &  4.75(18)   & 33.6(55)\0   \\
              & 4.997 &  0.6497 & 0.258(1) &  2.58(23)   & 25.0(24)\0   \\
              & 5.994 &  0.5229 & 0.507(2) &  1.18(15)   & 15.9(20)    \\

\hline
\hline
\end{tabular}
\end{center}
\end{table}

\setcounter{table}{0}
\begin{table}[t]
\noindent \caption{Continued.}
\bigskip
\begin{center}
\begin{tabular}
{lcc|ccc} \hline\hline
$L^*$=0.4 & $T^*$ & $\rho^*$ & $D^*$ & $\eta^*$ & $\lambda^*$  \\
\hline\hline

$\mu^{*2}$=0 & 1.899  & 0.5808 & 0.093(2)  & 2.51(20)   & 10.0(22) \\
             & 2.374  & 0.5123 & 0.187(1)  & 1.55(7)\0  & \07.8(11) \\
             & 2.848  & 0.4185 & 0.320(3)  & 0.91(9)\0  & \04.84(45) \\
\hline
$\mu^{*2}$=3 & 1.988  & 0.5881 & 0.088(1) & 2.97(15)    & 11.18(60) \\
             & 2.485  & 0.5179 & 0.182(1) & 1.80(18)    & \09.88(83) \\
             & 2.982  & 0.4262 & 0.351(2) & 0.94(5)\0   & \05.90(16)  \\
\hline
$\mu^{*2}$=6 & 2.156  &  0.5932 & 0.086(1) & 2.86(17)   & 11.4(12) \\
             & 2.695  &  0.5225 & 0.182(1) & 1.84(6)\0  & \09.95(90) \\
             & 3.234  &  0.4275 & 0.360(1) & 0.95(2)\0  & \06.22(60) \\
\hline
$\mu^{*2}$=9 & 2.347  & 0.5973  & 0.086(1) & 3.52(18)   & 14.70(57) \\
             & 2.934  & 0.5254  & 0.186(1) & 1.86(5)\0  & 10.88(24) \\
             & 3.521  & 0.4296  & 0.368(1) & 0.98(3)\0  & \08.02(60) \\
\hline
$\mu^{*2}$=12 & 2.571  & 0.5988  & 0.090(1) & 3.49(11)  & 13.4(15)\0 \\
             & 3.214  & 0.5243  & 0.194(1) & 1.92(19)   & 12.61(98) \\
             & 3.856  & 0.4257  & 0.390(1) & 1.16(8)\0  & \06.88(45) \\
\hline \hline
\end{tabular}
\end{center}
\end{table}

\setcounter{table}{0}
\begin{table}[t]
\noindent \caption{Continued.}
\bigskip
\begin{center}
\begin{tabular}
{lcc|ccc} \hline\hline
$L^*$=0.505 & $T^*$ & $\rho^*$ & $D^*$ & $\eta^*$ & $\lambda^*$  \\
\hline\hline

$\mu^{*2}$=0 & 1.642 &  0.5291 & 0.092(2)  & 2.24(20)   & \09.99(84) \\
             & 2.052 &  0.4637 & 0.184(7)  & 1.35(10)   & \06.01(60) \\
             & 2.463 &  0.3835 & 0.303(8)  & 0.83(7)\0  & \04.16(35) \\
\hline
$\mu^{*2}$=3 & 1.724 &  0.5346 & 0.087(1) & 2.66(20)   & \09.60(79) \\
             & 2.155 &  0.4711 & 0.175(1)  & 1.42(13)   & \08.10(94) \\
             & 2.586 &  0.3854 & 0.339(1)  & 0.85(2)\0  & \05.33(53) \\
\hline
$\mu^{*2}$=6 & 1.861 &  0.5392 & 0.082(1) & 2.92(6)\0  & 10.78(16)  \\
             & 2.327 &  0.4748 & 0.171(1)  & 1.44(2)\0  & \07.57(78) \\
             & 2.792 &  0.3902 & 0.334(2)  & 0.84(3)\0  & \05.09(45) \\
\hline
$\mu^{*2}$=9 & 2.027 &  0.5424 & 0.080(1) & 3.11(14)   & 10.81(69) \\
             & 2.535 &  0.4765 & 0.171(1)  & 1.67(11)   & \09.7(14)\0 \\
             & 3.041 &  0.3898 & 0.342(1)  & 0.88(3)\0  & \06.47(99) \\
\hline
$\mu^{*2}$=12 & 2.207 &  0.5457 & 0.079(1)& 3.38(6)\0  & 12.58(20)  \\
              & 2.759 &  0.4778 & 0.175(1) & 1.83(3)\0  & 10.2(16)\0 \\
              & 3.311 &  0.3886 & 0.351(2) & 0.83(20)   & \06.92(40) \\

\hline
\hline
\end{tabular}
\end{center}
\end{table}

\setcounter{table}{0}
\begin{table}[t]
\noindent \caption{Continued.}
\bigskip
\begin{center}
\begin{tabular}
{lcc|ccc} \hline\hline
$L^*$=0.6 & $T^*$ & $\rho^*$ & $D^*$ & $\eta^*$ & $\lambda^*$  \\
\hline\hline
$\mu^{*2}$=0 & 1.473 &  0.4900 & 0.091(2)  & 2.16(30)   & \09.24(46) \\
             & 1.842 &  0.4321 & 0.174(1)  & 1.14(1)\0  & \06.85(70) \\
             & 2.210 &  0.3520 & 0.295(9)  & 0.71(2)\0  & \04.15(40) \\
\hline
$\mu^{*2}$=3 & 1.536 & 0.4960 & 0.082(1) & 2.19(5)\0   & \07.95(31) \\
             & 1.920 & 0.4372 & 0.166(1)  & 1.27(7)\0   & \05.41(43) \\
             & 2.303 & 0.3597 & 0.317(1)  & 0.83(12)    & \04.78(40) \\
\hline
$\mu^{*2}$=6 & 1.665 &  0.4995 & 0.077(1) & 2.08(39)   & \09.8(18)\0 \\
             & 2.082 &  0.4398 & 0.160(1)  & 1.34(7)\0  & \07.73(40) \\
             & 2.498 &  0.3583 & 0.319(1)  & 0.79(10)   & \05.12(35) \\
\hline
$\mu^{*2}$=9 & 1.825 &  0.5006 & 0.076(1) & 2.91(12)   & 11.45(41) \\
             & 2.281 &  0.4378 & 0.165(1)  & 1.75(15)   & \07.00(95) \\
             & 2.738 &  0.3545 & 0.332(1)  & 0.84(1)\0  & \06.22(59) \\
\hline
$\mu^{*2}$=12 & 1.965 &  0.5062 & 0.071(1) & 3.28(12)   & \09.71(44) \\
              & 2.456 &  0.4429 & 0.159(1) & 1.62(4)\0  & \07.8(16)\0 \\
              & 2.948 &  0.3603 & 0.322(1) & 0.89(3)\0  & \05.94(40) \\
\hline \hline
\end{tabular}
\end{center}
\end{table}

\setcounter{table}{0}
\begin{table}[t]
\noindent \caption{Continued.}
\bigskip
\begin{center}
\begin{tabular}
{lcc|ccc} \hline\hline
$L^*$=0.8 & $T^*$ & $\rho^*$ & $D^*$ & $\eta^*$ & $\lambda^*$  \\
\hline\hline

$\mu^{*2}$=0 & 1.230 &  0.4302 & 0.082(2)  & 1.79(15)    & 8.28(42) \\
             & 1.538 &  0.3777 & 0.157(1)  & 1.11(14)    & 4.98(18)  \\
             & 1.845 &  0.3051 & 0.270(5)  & 0.60(4)\0   & 3.86(14) \\
\hline
$\mu^{*2}$=3 & 1.277 &  0.4364 & 0.071(1)  & 2.18(19)    & 8.52(31) \\
             & 1.596 &  0.3833 & 0.145(1)  & 1.27(14)    & 6.56(50) \\
             & 1.915 &  0.3101 & 0.287(1)  & 0.62(7)\0   & 3.67(20) \\
\hline
$\mu^{*2}$=6 & 1.390 &  0.4383 & 0.065(1)  & 2.25(45)   & 8.85(65) \\
             & 1.738 &  0.3835 & 0.141(1)  & 1.27(8)\0  & 5.7(11)\0 \\
             & 2.086 &  0.3085 & 0.288(1)  & 0.64(4)\0  & 3.39(60) \\
\hline
$\mu^{*2}$=9 & 1.513 &  0.4410 & 0.060(1)  & 2.44(10)   & 9.1(10)\0 \\
             & 1.891 &  0.3855 & 0.137(1)  & 1.39(1)\0  & 7.25(70) \\
             & 2.270 &  0.3093 & 0.286(2)  & 0.70(5)\0  & 4.32(34) \\
\hline
$\mu^{*2}$=12 & 1.629 &  0.4466 & 0.054(1)  & 3.02(8)\0  & 9.84(80) \\
             & 2.036 &  0.3899  & 0.130(1)  & 1.47(2)\0  & 7.52(26) \\
             & 2.443 &  0.3155  & 0.274(1)  & 0.74(5)\0  & 4.92(57) \\
\hline \hline
\end{tabular}
\end{center}
\end{table}

\setcounter{table}{0}
\begin{table}[t]
\noindent \caption{Continued.}
\bigskip
\begin{center}
\begin{tabular}
{lcc|ccc} \hline\hline
$L^*$=1 & $T^*$ & $\rho^*$ & $D^*$ & $\eta^*$ & $\lambda^*$  \\
\hline\hline
$\mu^{*2}$=0 & 1.058 &  0.3970 & 0.0640(5) & 1.71(16)   & 7.12(41) \\
             & 1.322 &  0.3487 & 0.128(1)  & 0.98(3)\0  & 4.75(35) \\
             & 1.587 &  0.2860 & 0.245(1)  & 0.62(6)\0  & 3.50(30) \\
\hline
$\mu^{*2}$=3 & 1.126 & 0.3976 & 0.0590(5)  & 1.96(3)\0   & 7.33(39) \\
             & 1.408 & 0.3463 & 0.127(1)   & 1.18(6)\0   & 4.73(13) \\
             & 1.689 & 0.2793 & 0.257(1)   & 0.56(4)\0   & 4.09(39) \\
\hline
$\mu^{*2}$=6 & 1.207 &  0.4041 & 0.0480(5) & 2.50(9)\0  & 7.45(78) \\
             & 1.508 &  0.3524 & 0.113(1)  & 1.24(20)   & 6.04(29) \\
             & 1.810 &  0.2887 & 0.232(1)  & 0.67(4)\0  & 3.62(17) \\
\hline
$\mu^{*2}$=9 & 1.308 &  0.4101 & 0.0390(5) & 3.31(13)   & 8.54(80) \\
             & 1.635 &  0.3576 & 0.104(1)  & 1.56(15)   & 5.22(18) \\
             & 1.961 &  0.2900 & 0.228(1)  & 0.72(5)\0  & 4.30(25) \\
\hline
$\mu^{*2}$=12 & 1.423 &  0.4163 & 0.0330(5) & 4.09(10)   & 8.98(30) \\
              & 1.779 &  0.3625 & 0.0970(5) & 1.75(5)\0  & 6.64(98) \\
              & 2.134 &  0.2923 & 0.223(1)  & 0.75(1)\0  & 3.92(34) \\
\hline \hline
\end{tabular}
\end{center}
\end{table}

\clearpage \listoffigures \clearpage

\end{document}